\documentclass[twocolumn]{revtex4}
\usepackage[dvips]{graphicx}
\usepackage{amsmath,amssymb}
\usepackage{pspicture, graphpap}

\newcommand\ltap{\
  \raise.3ex\hbox{$<$\kern-.75em\lower1ex\hbox{$\sim$}}\ }
\newcommand\gtap{\
  \raise.3ex\hbox{$>$\kern-.75em\lower1ex\hbox{$\sim$}}\ }

\newcommand\simge{\mathrel{%
   \rlap{\raise 0.511ex \hbox{$>$}}{\lower 0.511ex \hbox{$\sim$}}}}
\newcommand\simle{\mathrel{
   \rlap{\raise 0.511ex \hbox{$<$}}{\lower 0.511ex \hbox{$\sim$}}}}

\newcommand\CO{{\cal O}}

\newcommand\CZ{{\cal Z}}
\newcommand\be{\begin{equation}}
\newcommand\ee{\end{equation}}
\newcommand\bea{\begin{eqnarray}}
\newcommand\eea{\end{eqnarray}}
\newcommand\ba{\begin{array}}
\newcommand\ea{\end{array}}
\newcommand\nn{\nonumber}

\newcommand\mev{{\rm MeV}}
\newcommand\gev{{\rm GeV}}
\newcommand\tev{{\rm TeV}}

\newcommand\Gew{SU(2)_W\otimes U(1)_Y}

\begin{document}
\title{
\vskip -15mm
\begin{flushright}
\vskip -15mm
{\small BUHEP-06-03\\}
\vskip 5mm
\end{flushright}
Dark Matter in the Simplest Little Higgs Model}
\author{Adam Martin}
\email{aomartin@physics.bu.edu}
\affiliation{Department of Physics, Boston University \\
 590 Commonwealth Avenue, Boston, Massachusetts 02215}
\vspace{-1.0in}

\begin{abstract}
   We investigate a variation of the Simplest Little
   Higgs (SLH) model that is invariant under a discrete
   $\CZ_2$ symmetry. Our motivation for imposing this symmetry is the two dark
   matter candidates it implies, a scalar $\eta$ and a heavy neutrino. We
   examine the viability of these two candidates using the standard relic
   abundance calculation. Direct detection prospects are also discussed.
\end{abstract}

\maketitle

\pagebreak

\section*{I. Introduction}

The need for Dark Matter (DM), confirmed by current
astronomical observations, is a sufficient reason to look beyond the
Standard Model (SM). A reliable way to include DM into a model of physics beyond the Standard
 Model (SM) is to introduce a discrete symmetry under
which some new particles are odd, while all Standard Model (SM) particles are even. The
lightest odd particle is stable and is a candidate for DM. In supersymmetry (SUSY) this
discrete symmetry is $R$-parity, in Universal Extra Dimension (UED) models it is $KK$ parity, and for
Little Higgs models this discrete symmetry is called T
parity~\cite{Cheng:2003ju,Cheng:2004yc,Low:2004xc}. 

In this paper we will study the DM candidates of a $T$ parity invariant version
of the Simplest Little Higgs~\cite{Schmaltz:2004de, Kaplan:2003uc} model. Unlike other T
parity invariant Little Higgs models~\cite{Birkedal-Hansen:2003mp,
  Hubisz:2004ft}, it is not possible to make all new particles odd under T
parity. The gauge boson $Z'$ remains even. Since tree level $Z'$ exchange is allowed, the $Z'$ needs to be heavy
to avoid conflict with precision electroweak experiments. This means we must take the scale $f$ to be relatively
high. The price we pay for this is fine tuning. Despite this fine tuning, the
lightest $T$ parity odd field, which is either a scalar $\eta$ or a heavy neutrino
remains as an interesting DM candidate. The scalar
$\eta$ is similar to the electroweak singlet limit of the LH
scalar DM found in~\cite{Birkedal-Hansen:2003mp}. It annihilates
predominantly into higgs boson and $t \bar t $ pairs, but the cross section
is small unless $\eta$ is quite heavy. The heavy neutrino is a new
addition to Little Higgs model DM. It is also distinct from SUSY neutralino and
UED heavy neutrino DM since it is nearly inert under the SM weak gauge
group. The SLH heavy neutrino does couple to the heavy gauge bosons and to
the higgs. 

We investigate the viability of both of these DM candidates using the
standard relic abundance calculation. As an additional check, we also
explore the constraints direct detection experiments place on SLH
DM. In an effort to make this analysis more general, we do not limit ourselves to a
particular value of the symmetry breaking scale $f$. Hopefully this will allow
our analysis to be applicable to other models similar to SLH.

We find that $\eta$ can be an acceptable DM candidate only when coannihilating with a
slightly more massive fermion. The $\eta$ mass that yields the proper
abundance will depend on the scale $f$ and the type of fermion it
coannihilates with, but typically $m_{\eta} \sim 1.0\ \tev$. We find $\eta - nucleon$ elastic cross section to be well below
the current experimental sensitivity for the $\eta$ mass 
range of interest. The SLH heavy neutrino is an acceptable DM particle over a wider region of
parameter space. However, direct detection constraints
are also much stricter. Current direct
detection experiments rule out scenarios with $f \simle 3.0\ \tev$, and are
predicted to be sensitive to $f \sim 4.5\ \tev$ by the time
they finish running. To be consistent with the above bounds, heavy neutrino DM must have $m_N\ \sim 600\ \gev$ for $f = 3.0\ \tev$ and $m_N \sim 1.0\
\tev$ for $f = 4.5\ \tev$.

This paper is organized as follows: In section II we introduce the SLH model and show how it must be modified to
make it $\CZ_2$ invariant. We describe some of the existing constraints on
this model from precision electroweak observables in section III. After a brief overview of the standard relic abundance calculation in
section IV, we examine the DM candidates $\eta$ (section V) and the heavy neutrino (section VI) in
more detail. In section VII we include the effects of coannihilation, which
we find to be substantial. Section VIII is devoted to the experimental
constraints on SLH model DM from direct detection experiments.

\section*{II. Fields and Representations}
In the Simplest Little Higgs model, the SM weak gauge group is extended
to $SU(3)_w \otimes U(1)_x$. All SM electroweak doublets are therefore
accompanied by a new partner to form $SU(3)_w$ triplets. We follow the second
set of representations given in SLH since it has no $SU(3)^3_w$ anomaly. In this set of
representations, the
first and second generation quarks are promoted to a $\bar 3$ of $SU(3)_w$,
while the third generation quarks and the leptons
are a $3$. All
fields and their representations under $(SU(3)_c, SU(3)_w)_{U(1)_X}$ are listed below.
\bea
\Psi_{Q3} = (3,3)_{\frac{1}{3}} & & \Psi_{Q1},\Psi_{Q2} = (3,\bar 3)_0 ~~~~~
\Psi_{Li} = (1,3)_{\frac{-1}{3}} \nn \\
d^{c3} = (\bar3,1)_{\frac{1}{3}} & & d^{c1}_1,d^{c1}_2,d^{c2}_1, d^{c2}_2 =
(\bar 3,1)_{\frac{1}{3}} ~~~ e^c_i = (1,1)_{1} \nn \\
u^{c3}_1, u^{c3}_2 = (\bar 3, 1)_{\frac{-2}{3}} & & u^{c1}, u^{c2} = (\bar
3,1)_{\frac{-2}{3}} ~~~ n^c_i = (1,1)_0 \nn \\
& & ~~~~\Phi_{1}, \Phi_2 = (1,3)_{\frac{-1}{3}} \nn
\eea
 Under this assignment, the partner of the third generation SM
quark doublet is an up type quark $T$, while the first and second generation
partners are down type quarks $D,S$. The partner of the lepton doublet for
each generation $i$ 
is a neutrino $N_i$.

This extended weak gauge group is broken down to $\Gew$ when {\em two} different scalar $SU(3)$
triplet fields $\Phi_1, \Phi_2$ acquire vacuum expectation values.
\be
\langle \Phi_1 \rangle = \left(\begin{array}{c} 0 \\ 0 \\ f_1 \\
  \end{array}\right)~~,~~ \langle \Phi_2 \rangle = \left(\begin{array}{c} 0 \\ 0 \\ f_2 \\
  \end{array}\right)
\ee
Of the
10 pseudo-Goldstone boson (PGB) degrees of freedom in $\Phi_1\ {\rm and}\ \Phi_2$, 5 are eaten by
the gauge bosons of $(SU(3)_w \otimes U(1)_X)/(\Gew)$. The remaining 5
PGBs decompose under $\Gew$ as a complex doublet $h$, our higgs, and
a real scalar $\eta$.

The PGBs feel a potential generated radiatively by gauge and Yukawa interactions.  However, no single coupling breaks
enough of the global symmetry protecting the higgs to give it a mass. Terms
that include more than one coupling can give the higgs a mass, but these
terms are higher order in $\Phi_1, \Phi_2$ and the higgs mass terms they
generate depend only logarithmically
on the cutoff scale $\Lambda$. This 'collective breaking' is a necessary
ingredient in all Little Higgs models~\cite{Schmaltz:2005ky, Perelstein:2005ka}.  Along with gauge and
Yukawa interactions, this model also contains a tree level scalar potential $V =
\mu^2 \Phi^{\dag}_1 \Phi_2 + h.c.$. This potential gives a mass to $\eta$
and is needed to achieve the correct higgs vev with a higgs that sufficiently
heavy to avoid experimental bounds.  \\

{\em T parity operator $\hat{\Omega}$} \\

We now want to impose a $\CZ_2$ symmetry under which SM
particles are even and new particles are odd. The new particles predicted in
this model are the fermions $T,S,D,N_i$, the scalar $\eta$, and the
$SU(3)/SU(2)$ gauge bosons $W'_{\pm}, W'_{0,\bar 0}, Z'$.

We begin by looking at the
$SU(3)_w$ fermion triplets. We define the action of T parity on the
fermion triplets to be 
\be
\label{eq:tripTP}
\Psi_{Q,L} \rightarrow -\hat{\Omega}\Psi_{Q,L},
\ee
where $\hat{\Omega} = Diagonal(-1,-1,1)$. As desired, the first two
components of $\Psi_i$, a SM electroweak doublet, are even, while the
third component is odd. By defining (\ref{eq:tripTP}) with a minus sign, we have $\hat{\Omega} \in SU(3)$. This simplifies our T parity implementation
greatly. Since our Lagrangian is manifestly $SU(3)$
invariant, all factors of $\hat{\Omega}$ are
automatically taken care of. All that remains to be determined is 
a $\pm$ sign for each field. We can determine the sign by whether the field is
a SM field or a new field. For the higgs triplets, the transformation is
\be
\Phi_1 \rightarrow +\hat{\Omega}\Phi_2.
\ee
which requires $f_2 = f_1 = f/\sqrt2$. Under this transformation, the higgs
doublets are T parity even, while the remaining PGB, $\eta$, is T parity
odd. The vevs of the $\Phi$ are also T parity even.

Having defined the action of T parity on $SU(3)$ triplets, we immediately know how
the octet of gauge fields $A_{\mu}$ transform.
\be
A_{\mu} \rightarrow \hat{\Omega}A_{\mu}\hat{\Omega}
\ee
This transformation leaves the diagonal gauge bosons and the gauge bosons in the upper left
2 by 2 block even, while all other gauge bosons are odd. This is immediately a problem since
the $Z'$, a new particle, remains even.  For now we
press on with our discrete transformation, but we will
discuss the consequences of an even $Z'$ in more detail in the next section. \\

{\em Enforce T parity on the Yukawa sector:} \\

To determine how the rest of the fields transform, we need to
enforce T parity invariance on the Yukawa sector. To be consistent with the T
parity transformation of the higgs, this means that both $\Phi_1\ {\rm
  and}\ \Phi_2$
must couple with equal strength.

For the first two generations of up quarks, the bottom quark, and the leptons,
enforcing T parity is straight forward and requires no additional operators. The fields $d^{c3}, u^{c1}, u^{c2},
{\rm and}\ e^{c}_i$ are all T parity even. Enforcing T parity on the down quark
Yukawas is slightly more difficult, for reasons that become clear by
examining the top Yukawa in greater detail.

The
operator that leads to the top mass is 
\be
\label{eq:topyuk1}
\lambda_{u}\Psi_{Q3}(\Phi^{\dag}_1 u^{c3}_1 + \Phi^{\dag}_2 u^{c3}_2).
\ee

The transformation properties of the $u^{c3}_i$ are determined by requiring
that the linear combination of the $u^{c3}_i$ that gets a mass with the heavy
top quark is odd. Expanding out the higgs fields to lowest order, we find
\be
\label{eq:topyuk2}
-i\frac{\lambda_u h}{2}(u^{c3}_1 - u^{c3}_2)Q_3 + \lambda_u\frac{f}{\sqrt 2} (u^{c3}_1
+ u^{c3}_2)T + \cdots .
\ee
The correct T parity transformation for the $u^{c3}_i$ is therefore
\be
u^{c3}_1 \rightarrow - u^{c3}_2.
\ee
Comparing (\ref{eq:topyuk1}) to the corresponding equation in SLH, we see
that we have halved the number of parameters: $f_1, f_2,
\lambda_{u1},\lambda_{u2} \rightarrow \lambda_u, f$. The reduced number of
parameters leads us to a simple relation between the mass
    of the top quark and the mass of new heavy quark $T$,
\be
\label{eq:TPrat}
\frac{m_{T}}{m_{t}}  = \Big(\frac{\sqrt 2f}{v}\Big).
\ee
For $f = 2.0\ \tev$ this ratio gives $m_T = 2.0\ \tev$, almost twice its
value at the 'golden point' of SLH.

 The problem with the first and second
generation down type Yukawas in SLH is now clear. The operator
\be
\lambda_d \Psi_{Q1}(\Phi_1 d^{c1}_1 + \Phi_2 d^{c1}_2)
\ee 
has exactly the same form as (\ref{eq:topyuk1}), and would imply a
relationship similar to (\ref{eq:TPrat}) between the SM down quark mass and
the heavy $D$ quark mass. For $m_d \approx 10\ \mev$ This relation predicts a heavy down quark with a mass less than a GeV! Clearly we need to use
a different operator to give mass to the heavy down type quarks.

To avoid this catastrophe, we introduce operators that give mass to the T
parity even and T parity odd
portions separately
\be
\label{eq:DTYnewops}
 \lambda_d \Psi_{Q1}(\Phi_1 + \Phi_2)d^{c1}_H + \epsilon_d
 \Psi_{Q1}(\Phi_1 - \Phi_2)d^{c1}_L,
\ee
where $ d^{c1}_H = (d^{c1}_1 + d^{c1}_2),\ d^{c1}_L = (d^{c1}_1 - d^{c1}_2)$
are respectively the odd and even combinations~\footnote{The choice of which combination of $d^{ci}_a$
  to be odd/even here is arbitrary. We chose these particular combinations to be consistent
  with the $u^{c3}_i$.} of $d^{ci}_1, d^{ci}_2$. 
The new particles and SM quark masses are no longer related for $\Psi_{Q1,2}$.

The only remaining term is the mass term for the heavy neutrinos
$N_i, n^c_i$, both of which should be T parity odd. The operator
\be
\label{eq:neutrinoyuk}
\lambda_n n^c(\Phi^{\dag}_1 + \Phi^{\dag}_2)\Psi_L
\ee 
is sufficient. 

The couplings (\ref{eq:DTYnewops},\ref{eq:neutrinoyuk}) violate the little
higgs mechanism. By this we mean that
they generate the operator $\Phi_1^{\dag}\Phi_2 + h.c.$  at one loop, and
therefore lead to quadratically divergent radiative higgs mass terms
\be
\label{eq:deltamhsq}
\delta m^2_h \sim \lambda^2 f^2.
\ee
The larger the mass of the heavy quark or neutrino, the larger the
contribution to the higgs mass. Since we are not aiming to remove fine
tuning in this model, we will not restrict particle masses on account of
(\ref{eq:deltamhsq}). However it does encourage us to look for light $\simle
1.0\ \tev$ DM.

Now that we have made the SLH model consistent with T parity we can identify DM candidates. The DM
particle will be the lightest T parity odd particle (LTOP). The neutral T
parity odd fields available are
the neutrinos $(N_i,n^c_i)$, the scalar $\eta$, and the heavy gauge bosons $W'_{0, \bar 0}$. The mass
of the heavy neutrino can easily be a few
hundred GeV for a small value of $\lambda_n$, and we expect the mass of 
$\eta$ ($\sim \mu$) to be approximately the weak scale. The
$W'_{0,\bar 0}$ obtain a mass $m_{W'} = \frac{g f}{\sqrt 2}$ when $SU(3)_w
\otimes U(1)_X$ breaks to $\Gew$. The $W'$ mass and all of its interactions are fixed for a given scale $f$. Because of this lack of flexibility and
because the $W'$ is usually heavier than the $\eta$ and heavy neutrino, we
will not consider the $W'$ as a DM
candidate here.

\section*{III. Parameters and Tunings}
  
 As we are unable to make the $Z'$ odd under T parity, we need it to be very heavy in
 order to avoid conflict with precision electroweak observables (EWPO). The
 only way to get a heavy $Z'$ in the SLH model is to increase the symmetry breaking
 scale $f$. The lower bound on $f$ was originally estimated to be $f \simge
 2.0\ \tev$~\cite{Schmaltz:2004de}, but later analysis point to a stricter bound of $f \simge 4.5\
 \tev$~\cite{Han:2005dz, Marandella:2005wd, Casas:2005ev}. Rather than abide by
 one of these bounds, we will investigate SLH DM for a variable scale $f \ge
 2.0\ \tev$. This allows our analysis to be more general, and applicable to other similar
 models. Whenever results for a fixed $f$ are desired, we will use the value
 $f = 4.0\ \tev$.

For a given $f$, the masses of the heavy gauge bosons and heavy top are automatically determined.
\bea
\label{eq:NPmass}
m_T &=& m_t \Big( \frac{\sqrt 2 f}{v} \Big),~~ M_{Z'} = \frac{\sqrt 2 g f}{\sqrt{3
  - \tan{\theta_W}^2}},~~M_{W'} = \frac{g f}{\sqrt 2} \nn \\
\eea
The radiative contributions of these heavy particles to the higgs mass can
then be determined by the usual technique as shown
in~\cite{Schmaltz:2004de,Kaplan:2003uc}. As a consequence of a high value of $f$, the
masses in (\ref{eq:NPmass}) can be quite large, causing the theory to be somewhat finely tuned. However, since it is not our
goal in this paper to remove fine tuning, we will accept it at its current
level.

For relatively light DM ($\simle 1\ \tev$), we must also check to make sure that
any effective four fermion operators are appropriately suppressed. Four
fermions operators come about in this scheme through box diagrams with SM external legs and T
parity odd particles on the internal loop. The constraints on new
contributions to four electron operators $(eeee)$ are the most stringent,
followed by constraints on four quark operators $(qqqq)$. We have estimated
the contributions of heavy particle loops to $(eeee), (qqqq)$ in the model we have presented, and we find that these
contributions are less than $1/(5 - 10\ \tev)^2$ in the DM mass region of interest. The $(qqqq)$ contributions from $D-\eta$ boxes are suppressed by factors of
$(v/f)$ at each vertex (basically we have decoupled the heavy quarks from
their SM partners by introducing the extra operators). Charge conservation
forbids a $e N \eta$ vertex, so only $W' - N$ box diagrams contribute
to $(eeee)$. These contributions are safely suppressed by the larger mass of the $W'$.

\section*{IV. Relic DM Abundance Calculation}

We now give a brief overview of the relic DM
calculation before examining the heavy neutrino and $\eta$ further. We
follow the standard procedure outlined in~\cite{Srednicki:1988ce, Griest:1988ma,
  Servant:2002aq}.

 The number density $n$ of a cold dark matter particle $\chi$
obeys the Boltzmann equation
\be
\label{eq:Boltz}
\frac{dn}{dt} + 3Hn = -\langle \sigma v_{rel}\rangle(n^2 - n^2_{eq}).
\ee
    
Here $H$ is the expansion rate of the universe, $n_{eq}$ is the equilibrium
number density, and $\sigma v_{rel}$ is the annihilation cross section of the
particle times
the relative velocity. The $\langle~\rangle$ around the annihilation
cross section indicates that we take a
thermal average. We will use the standard value for the equilibrium
density of a cold, nonrelativistic particle of mass $m$ with $g$ degrees
of freedom:
\be
n_{eq} = g\Big(\frac{mT}{2\pi}\Big)^{3/2}exp(-m/T).
\ee

Since we expect the cold DM particles to be slow, we first take the
non-relativistic limit of the cross section. To obtain the nonrelativistic cross section,
we substitute $s = 4M^2_{\chi} + M^2_{\chi} v^2$ into the cross section and
keep only
first order terms in $v^2$. Thermal averaging then gives
\be 
\label{eq:sigmavrel}
\sigma v_{rel} \approx_{NR} a + b v^2 \rightarrow \langle \sigma v_{rel}
\rangle \approx a + \frac{6b}{x},
\ee
where $x = \frac{M_{\chi}}{T}$.

From the thermally averaged cross section we can determine the contribution of
particle $\chi$ to the total energy density of the universe. The density of particle $\chi$
divided by the critical density corresponding to a flat universe, denoted as
$\Omega_{\chi} h^2$, is determined to be
\be
\label{eq:Econt}
\Omega_{\chi} h^2 \approx \frac{1.04\times10^9}{M_{pl}}\frac{x_F}{\sqrt{g^*}(a + \frac{3b}{x_F})}.
\ee
 Here $x_F$ is the critical temperature below which the expansion term alone determines
the evolution of the number density. It is referred to as the
freezeout temperature and it can be expressed analytically as 
\be
\label{eq:freeze}
x_F = \log{\Big{(}\frac{0.047~g~M_{\chi}~M_{pl}~(a + \frac{6b}{x_F})}{\sqrt{g^* x_F}} \Big{)}}.
\ee
The parameter $g^*$ is the total (spin, color etc.)
number of relativistic degrees of freedom at temperature
$x_F$. Typically, $x_F \approx 20$.

The results (\ref{eq:Econt},\ref{eq:freeze}) are derived by solving the Boltzmann equation assuming
constant entropy per comoving volume in the limits $x \ll x_F$ and $x \gg
x_F$, then matching the
two solutions together. A more complete discussion of this procedure can be
found in~\cite{Srednicki:1988ce, Servant:2002aq}. We will consider DM to be
cosmologically acceptable if $\Omega_{\chi} h^2$ falls within the $2\ \sigma$ limits from WMAP, $0.094 \le \Omega_{\chi} h^2 \le 0.129 $~\cite{Spergel:2003cb,Bennett:2003bz}. \\

{\em Coannihilation}\\

If there is a particle ($\chi_2$) just slightly heavier than the LTOP
($\chi_1$), then coannihilation processes $\chi_1 \bar{\chi_2}\rightarrow X
X'$, etc. become important in determining the LTOP number
density~\cite{Griest:1990kh}. To study the effects of coannihilation we must modify our relic abundance
formalism slightly.  The formalism of (\ref{eq:Econt}, \ref{eq:freeze})
assumes that all T parity odd particles have decayed into the
LTOP by $x_F$. In order to consider particles with approximately the same
mass as the LTOP we must drop this assumption and examine the evolution of
 the total number density of T parity odd particles $n_{\chi} = n_{\chi_1} +
n_{\chi_2} + \cdots$. The evolution equation for the total number density
takes the familiar form
\be
\label{eq:coanboltz}
\frac{dn}{dt} + 3 H n = -\langle \sigma_{eff} v_{rel}\rangle (n^2 - n^2_{eq}).
\ee
Here $\sigma_{eff}$ is the sum of all cross sections $\sigma_{ij} = \sigma(\chi_i \chi_j
\rightarrow XX')$ weighted by the degrees of freedom $g_i$ in $\chi_i$ and the mass
difference $\Delta_i = (m_{\chi_i} - m_{\chi_1})/m_{\chi_1}$. For $N$ coannihilating
particles, we have
\be
\label{eq:sigmaeff}
\sigma_{eff} = \sum_{ij}^N \sigma_{ij}\frac{g_i g_j}{g^2_{eff}}(1 +
\Delta_i)^{3/2}(1 + \Delta_j)^{3/2}e^{-x(\Delta_i + \Delta_j)},
\ee
where
\be
g_{eff} = \sum_i g_i (1 + \Delta_i)^{3/2}e^{-x \Delta_i}.
\ee
Note that the mass appearing in $x = \frac{m_{\chi_1}}{T}$ above is the LTOP mass.
The effects of particles much heavier than the LTOP are exponentially
suppressed, so only particles very close in mass to the LTOP
are relevant in the sum (\ref{eq:sigmaeff}).

The solution to (\ref{eq:coanboltz}) is the same as
(\ref{eq:Econt},\ref{eq:freeze}) but with modified $a$ and $b$.
\be
 \Omega_{\chi_i} h^2 \approx \frac{1.04\times10^9}{M_{pl}}\frac{x_F}{\sqrt{g^*}(I_a + \frac{3I_b}{x_F})},
\ee
where
\be
I_a = x_F\int_{x_F}^{\infty}\frac{a_{eff}}{x^2}~{\mathrm dx}\ ,\  I_b = 2x^2_F
\int_{x_F}^{\infty}\frac{b_{eff}}{x^3}~{\mathrm dx} \nn \\
\ee
The averaged coefficients $a_{eff}, b_{eff}$ are defined by substituting $a_{ij},b_{ij}$ for $\sigma_{ij}$ in (\ref{eq:sigmaeff}).

We now examine the relic abundance calculations for
$\eta$ and $N$ in more detail to locate the regions in parameter space
where they are suitable DM candidates. For a given value of $f$, we will vary  $\mu, \lambda_n, {\rm and}\ \lambda_d$ (or equivalently
$m_{\eta}, m_N, m_D$). For simplicity we will first
consider the scenario where the LTOP is much lighter than the other new
particles. Once we understand the abundances in the $m_{LTOP} \ll m_{NP}$ limit, we will explore
the effects of coannihilation. 

\section*{V. Scalar $\eta$ as the LTOP}

The first dark matter candidate that we consider is $\eta$, the remaining
PGB from $SU(3)_w \otimes U(1)_X \rightarrow \Gew$ breaking. It
receives a mass $m^2_{\eta} \sim
\mu^2$ from the scalar potential.

To determine the viability of $\eta$ dark matter we need to
calculate the annihilation cross section. As $\eta$ is an EW
singlet, it does not couple directly to the $W^{\pm}$ or $Z^0$ gauge
bosons.  The $\eta \eta Z'$ couplings
is also zero because $\eta$ is a real scalar field. The dominant annihilation mode for $m_{\eta} \simle f$ is into a pair of
higgs bosons $\eta\eta \rightarrow h_0 h_0$. This annihilation proceeds
through the $\eta^2 h^{\dag} h$ term in the scalar potential, and also
through the $t$-channel exchange of a heavy $W'_0$ gauge boson.

The other important annihilation mode for $\eta$ is $\eta \eta \rightarrow t \bar t$.
As the $\eta T t^c$ coupling is large, we might expect $\eta \eta \rightarrow
t \bar t$ through $T$ exchange to be the dominant annihilation mode. However, this
process is helicity suppressed, meaning that this process has $a \cong 0$ 
(see Eq. \ref{eq:sigmavrel}). The $b$ term for $\eta \eta
\rightarrow t \bar t$ can be large, but its effect on the overall cross
section is suppressed by an additional factor of $x_F \sim \CO(20)$.

There is also a contribution to $\eta\eta \rightarrow t \bar t$ from higgs
exchange.  The
interaction responsible for this annihilation comes from the scalar potential
and has strength $\sim
\frac{m^2_{\eta}}{f^2}$. For heavy $\eta$ this becomes $\CO(1)$. Annihilation
modes of $\eta$ to light fermions or SM gauge bosons are all negligible,
suppressed by small fermion masses or by additional powers of $(v/f)$.

In Figure 1. we plot the regions where $\Omega_{\eta} h^2$ is allowed by
cosmology as a function of the mass of $\eta$ and $F$. We assume a higgs mass~\footnote{Calculations with $m_h = 115\ \gev\ {\rm
    and}\ m_h = 200\ \gev$ yield almost identical results.} of $140\
\gev$.
\begin{figure}[!h]
  \begin{center}
    \includegraphics[width=2.5in,height=3.5in, angle=270]{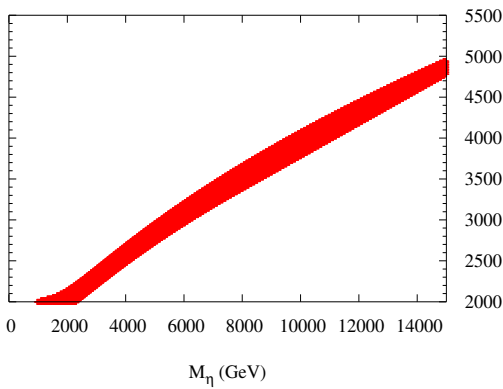}
    \caption{Regions in the $m_{\eta} - f$ plane where $\Omega_{\eta}
      h^2$ is within $2\ \sigma$ of the WMAP limits. In this scenario we assume all
      other T parity odd particles are much heavier than $\eta$.}
    \label{fig:relic_eta}
  \end{center}
\end{figure}
For a given $f$,  we see that $\Omega h^2_{\eta}$ does not fall into the allowed
range until the $\eta$ mass is $\CO(f)$. Because of the higgs exchange contribution, $\eta$
annihilation into $t \bar t$ continues to grow as $m_{\eta}$ is increased. When $m_{\eta} \sim f$,
the cross section is elevated to the point that we get the right relic
abundance. Once $m_{\eta}$ is much larger than $f$ the cross section is
too big and $\eta$ DM is insufficient. Although there is a $m_{\eta} - f$
region where $\Omega_{\eta} h^2$ is 
within the WMAP experimental limits, we cannot accept it as a viable DM
candidate. The $\eta$ mass which gives the right relic abundance is {\em always}
heavier than the mass of the $W'$, and thus the $\eta$ in this scenario would
never be the LTOP. As we will see in section VII., the acceptable
$m_{\eta} - f$ region changes significantly when we allow coannihilation with nearly
degenerate fermion.

\section*{VI. Heavy Neutrino(s) as the LTOP}

The heavy neutrino is a slightly more complicated LTOP since it is a dirac
fermion. To calculate its annihilation cross section we must
calculate the annihilation cross sections of each of it's Weyl components $(N, n^c)$
with themselves and with each other.
\bea
\label{eq:diracN}
\sigma_{tot} &=& \sigma(N N \rightarrow X X') + \sigma(n^c n^c \rightarrow X X')
+ \nn \\
& & ~~~\sigma(N \bar N \rightarrow X X') + \sigma(n^c \bar n^c \rightarrow
X X') \nn \\ 
& & ~ + \sigma(N n^c \rightarrow X X')  + \sigma(N \bar n^c
\rightarrow X X') + h.c., \nn \\
\eea
where $X, X'$ are some SM particles. Though this looks tedious, many of the
sub-cross sections are either zero or very small. The only significant terms
are $\sigma(N \bar N \rightarrow X X')$ and $\sigma(N n^c
\rightarrow X X')$.

 Unlike many heavy neutrino DM candidates~\cite{Servant:2002aq,
   Jungman:1995df}, the SLH heavy neutrino interacts very weakly with
 SM gauge bosons. It does not couple to the $W^{\pm}$, and its coupling to the $Z^0$ is suppressed by a
factor of $\frac{v^2}{8f^2}(1 - \tan^2{\theta_W})$. All annihilations proceed
through the exchange of a heavy gauge boson, a heavy fermion, or the higgs.

 The $N \bar N$ annihilation cross section contains the mode $N \bar N \rightarrow \nu \bar{\nu},
\ell^-\ell^+$ through $t$-channel $W'$ exchange. For light
$m_N$ this is the dominant
mode as the coupling of the $N$ to the $W'$ is
large. For heavier $m_N$, $N \bar N$ annihilation to fermions via $s$-channel $Z'$ quickly becomes the most
important process. The reason for this dominance is simply the presence
of a pole at $m_N = M_{Z'}/2$ in the $s$-channel propagator. In vicinity of
$m_N = M_{Z'}/2$ the cross section is so large that its nonrelativistic limit
must be treated specially. The usual Taylor expansion about $v = 0$ (Eq. (\ref{eq:sigmavrel})) is
very inaccurate and can result in negative cross
sections\footnote{Strictly speaking, this inaccuracy is problematic only for $m_N >
  \Gamma_{Z'}$. We calculate $ \Gamma_{Z'} \approx 70\
  \gev \ll m_N$. }. To avoid this inaccuracy, we follow
one of the prescriptions suggested in~\cite{Griest:1990kh} and set $v = 0$ in
all $s$ channel $Z'$ propagators.

 For $m_N \gg M_{Z'}/2$,
$N n^c$ annihilation into heavy SM quarks via higgs exchange may also become relevant as the $h N
n^c$ coupling is proportional to $\frac{m_N v}{f^2}$. It is interesting
to note that, contrary to many DM candidates~\cite{Servant:2002aq,Cheng:2002ej,Birkedal-Hansen:2003mp, Hubisz:2004ft,Mahbubani:2005pt}, the annihilation into
$W^{+}W^{-}$ is not very significant for either DM candidate considered here.

We have calculated the relevant terms in (\ref{eq:diracN}) and the resulting
$\Omega_N h^2$ is plotted in Figure 2. for both fixed $f$ and variable $f$. \\ 
\begin{figure}[!ht]
  \begin{center}
    \includegraphics[width=3.75in,height=2.85in, angle=270]{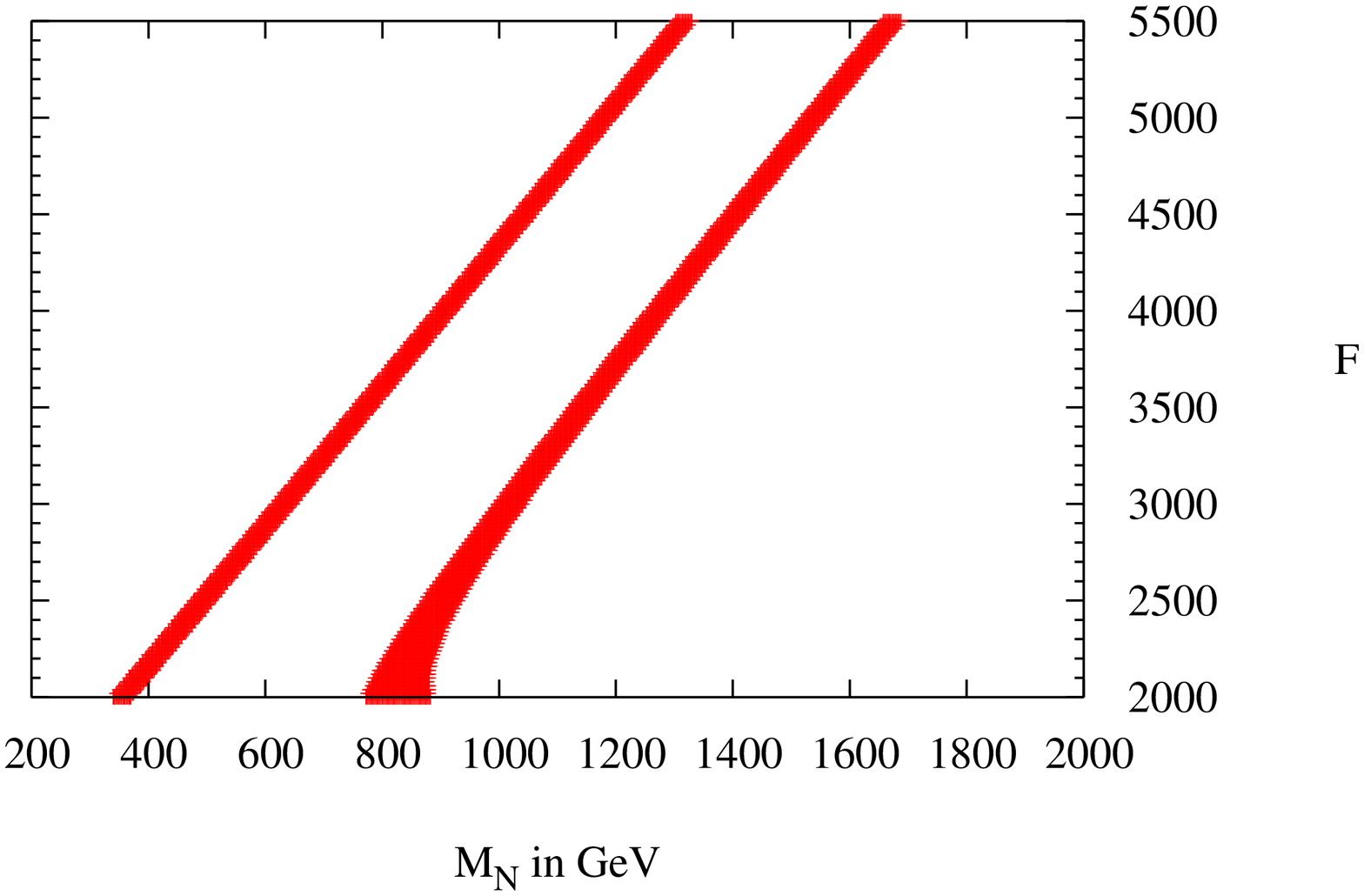}
    \includegraphics[width=2.3in,height=3.0in]{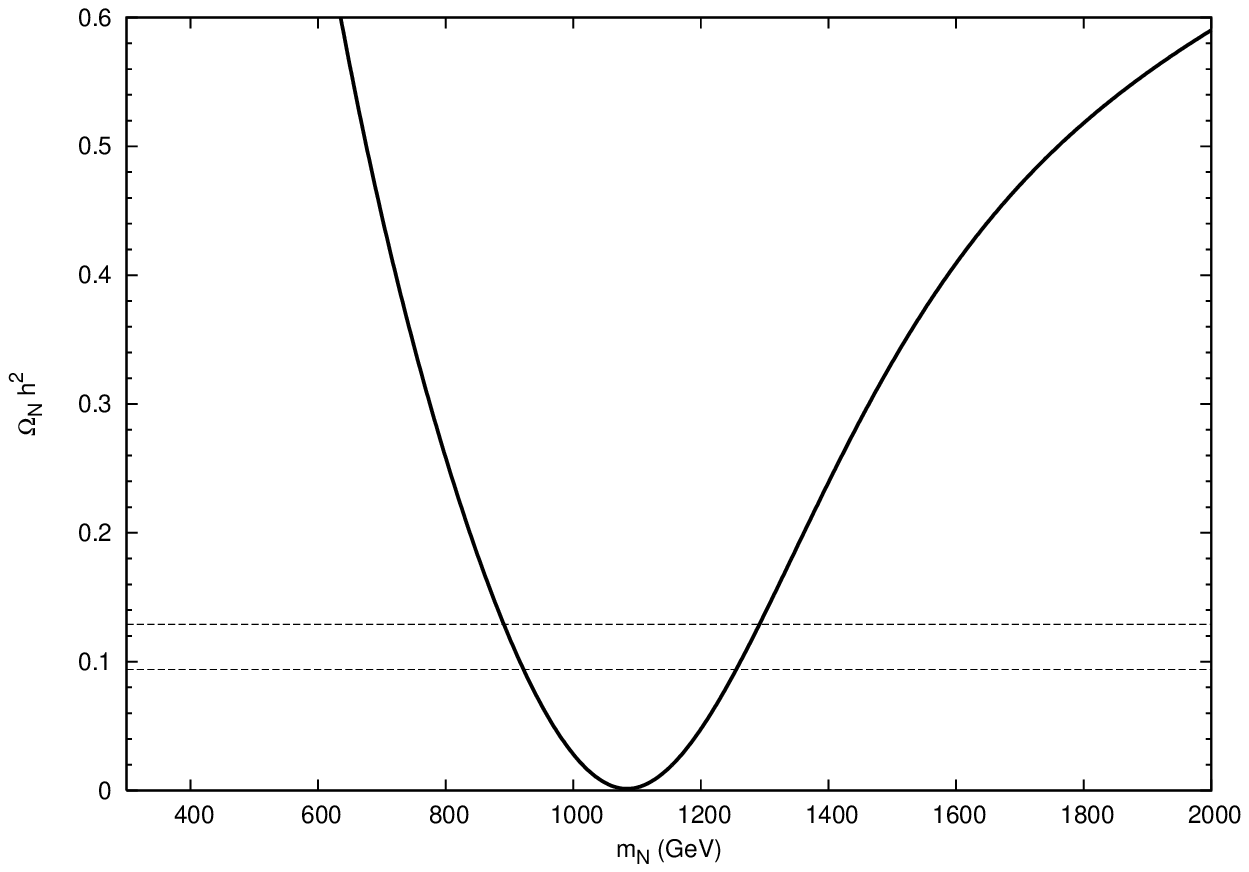}
    \caption{In the top plot: We
      show the regions where $\Omega_N h^2$ is within the $2\ \sigma$ WMAP
      limits for a range of $f$ and $m_N$. In the bottom plot: $\Omega_N h^2$ for a heavy neutrino LTOP versus the mass of the
      heavy neutrino in $\gev$ for fixed $f = 4.0\ \tev$. In both plots we have assumed that
      all other T parity odd particles have mass $\gg m_N$.}
    \label{fig:relic_N}
  \end{center}
\end{figure}
In the $m_N - f$ plane, we see two strips where the $\Omega_N h^2$ is
consistent with experiment. The $m_N$ values of the both strips increases
with $f$, yet they remain roughly $400\ \gev$ apart. Between the two strips, heavy
neutrino DM is unacceptable because there is not enough of it, while outside
the strips there is too much DM.  We can understand the origin of these two
regions by looking at the relic abundance for fixed $f$
as a function of the heavy neutrino mass, shown in the rightmost plot
above. Right at the $Z'$ pole, the annihilation cross
section blows up causing the relic abundance to plummet. However, as $m_N$
moves away from the resonant value $M_{Z'}/2$ by $\sim \pm 200\ \gev$, $\Omega_N h^2$ crosses the allowed WMAP region. The acceptable $m_N$ regions always
border the $Z'$ pole. Thus as $f$ is increased, raising $M_{Z'}$, the allowed
neutrino DM mass also increases.

\section*{VII. Including Coannihilation}

As we saw in section IV, particles slightly heavier than the LTOP can
have a significant effect on the relic DM abundance. The nature of this effect
depends on the self-annihilation cross section and the number of
degrees of freedom of the heavy particle, as well the
annihilation cross section of the heavier particle with the LTOP. Coannihilation of the LTOP with a
particle that has the same (or fewer) degrees of freedom and participates in
the same interactions usually results in a smaller cross section, while
coannihilation with a strongly interacting particle with many degrees of
freedom can increase the cross section by an order of magnitude or more. In
the Simplest Little Higgs model there can be 
coannihilation among more than one generation of heavy neutrino, between the heavy
neutrinos and the heavy quarks, and also
between $\eta$ and a heavy fermion\footnote{There can also be coannihilation
  of either DM candidate with the $W'$ but we have neglected these effects here.}. \\

{\em Heavy neutrino LTOP including coannihilation} \\

We first consider coannihilation among three degenerate
heavy neutrino flavors. Two neutrinos of the same flavor can annihilate with
each other as
described previously, and two neutrinos of different flavor can annihilate
through the $t$-channel exchange of a $W'$. The flavor changing processes $N_i \bar N_j
\rightarrow \nu_i \bar{\nu}_j, \ell^-_i \ell^+_j$ are less efficient than the
self-annihilation processes, and as a result the average cross section at a given mass $m_N$ for three neutrino flavors is
smaller than the cross section for a single neutrino flavor. This results in a
larger $\Omega_N h^2$, as can be seen in Figure 3. \\ 
\begin{figure}[!h]
  \begin{center}
    \includegraphics[width=3.75in,height=2.85in, angle=270]{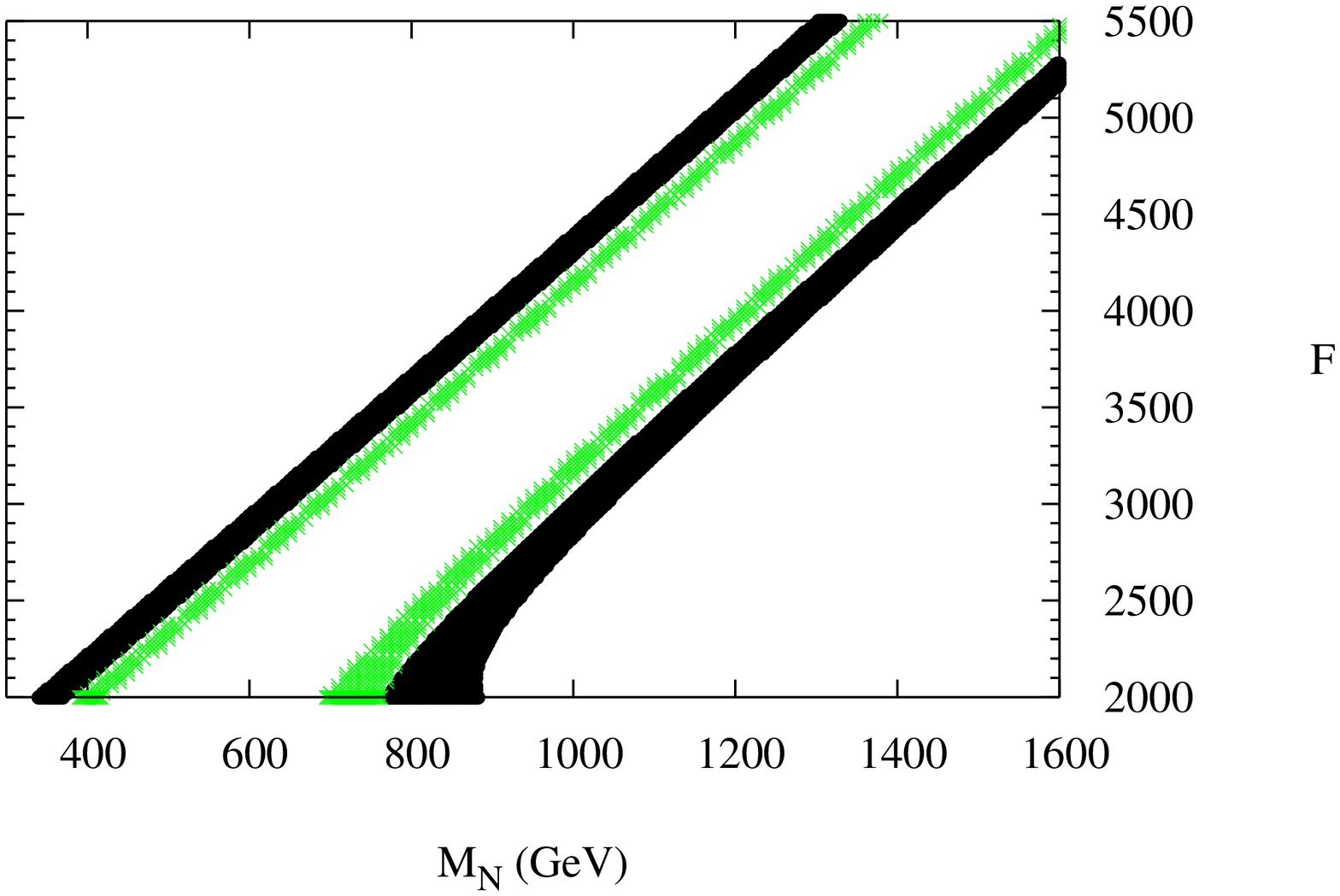}
    \includegraphics[width=2.3in, height=3.0in]{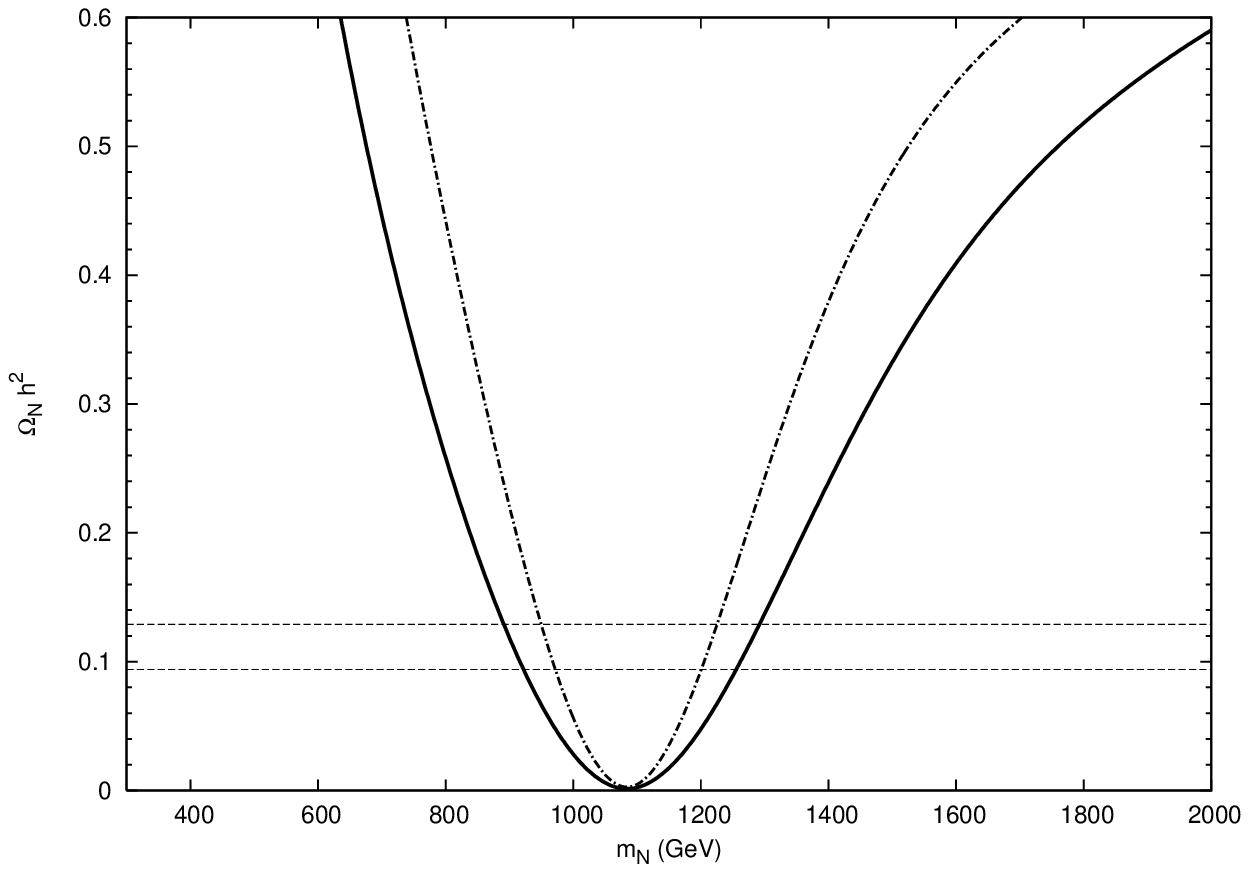}
    \caption{Top: Acceptable $\Omega_N h^2$ for a single flavor of heavy neutrino (black) and
      including coannihilation among three flavors of heavy neutrino (gray,
      green online)
      versus $m_N$, $f$. All neutrinos are degenerate. Right: $\Omega_N h^2$
      for one (solid line) and three
      (dash-dot line) degenerate neutrino flavors for $f = 4.0\ \tev$. Dashed
      lines are WMAP limits.}
    \label{fig:relic_coN}
  \end{center}
\end{figure}
In order to achieve the correct relic abundance with three neutrinos, we
need a larger self annihilation cross section to compensate for the extra
degrees of freedom. This requires $m_N$ to be even closer to $M_{Z'}/2$, and thus
the two acceptable DM regions are closer together. 

Coannihilation of a heavy neutrino with a heavy quark has a very different effect. Since
they are colored, heavy quarks can annihilate to SM quarks through gluon
exchange, and thus $\sigma(D\bar D \rightarrow XX')\ {\rm and}\
\sigma(d_H^c \bar d^c_H \rightarrow X X')$ include $\CO(\alpha^2_s)$
terms.  The quark self-annihilation cross sections are consequently orders of magnitude larger than $\sigma(N\bar N
\rightarrow XX')$, especially for small $m_N, m_D$. As the heavy quarks couple to the
$Z'$, their cross section is also enhanced near the pole at $m_D =M_{Z'}/2$.

The large heavy quark self annihilation cross sections dominate
the average cross section $\sigma_{eff}$ since they are weighted by the large number of degrees
of freedom in the colored heavy quarks.  The mixed annihilation channel $\sigma(N \bar D \rightarrow XX')$ through
$W'$ exchange is enhanced by color factors relative to the mixed neutrino
annihilation cross section, but it is still much smaller than the self
annihilation cross sections $\sigma(N\bar N \rightarrow XX'), \sigma(D\bar D
\rightarrow X X')$. However, because $\sigma_{eff}$ is so strongly controlled by
$\sigma(D\bar D \rightarrow XX')\ {\rm and}\
\sigma(d_H^c \bar d^c_H \rightarrow X X')$, the
inefficiency of the mixed $N\bar D$ processes
has a negligible effect on the average cross section except for when $m_D,
m_N$ are very close to $m_{Z'}/2$.

In Figure 4. we plot the acceptable regions of heavy neutrino DM including coannihilation
with a heavy quark as a function of $m_N, f$.
\begin{figure}[!ht]
  \begin{center}
    \includegraphics[width=3.75in,height=2.85in, angle=270]{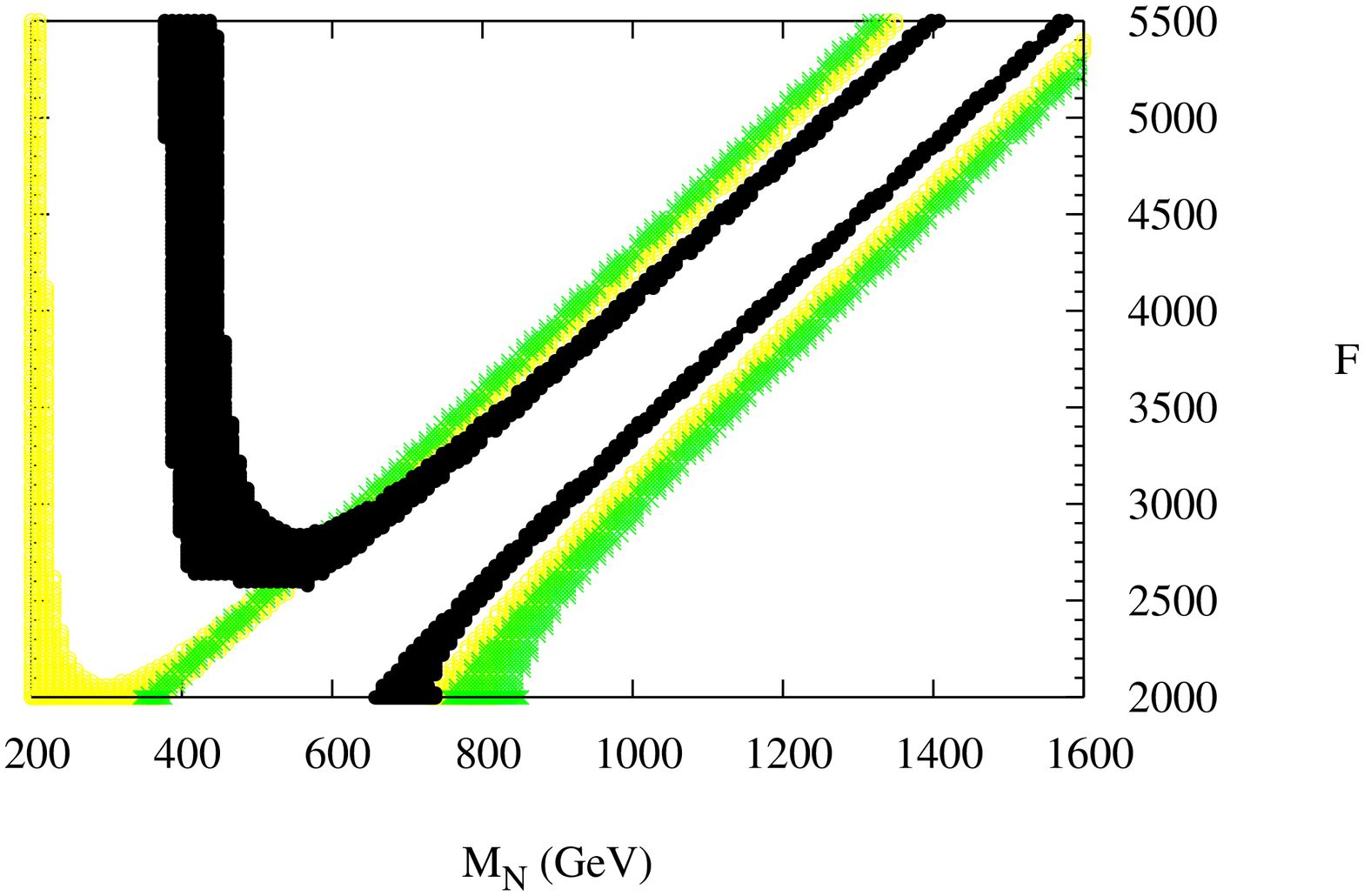}
        \includegraphics[width=2.3in,height=3.0in,]{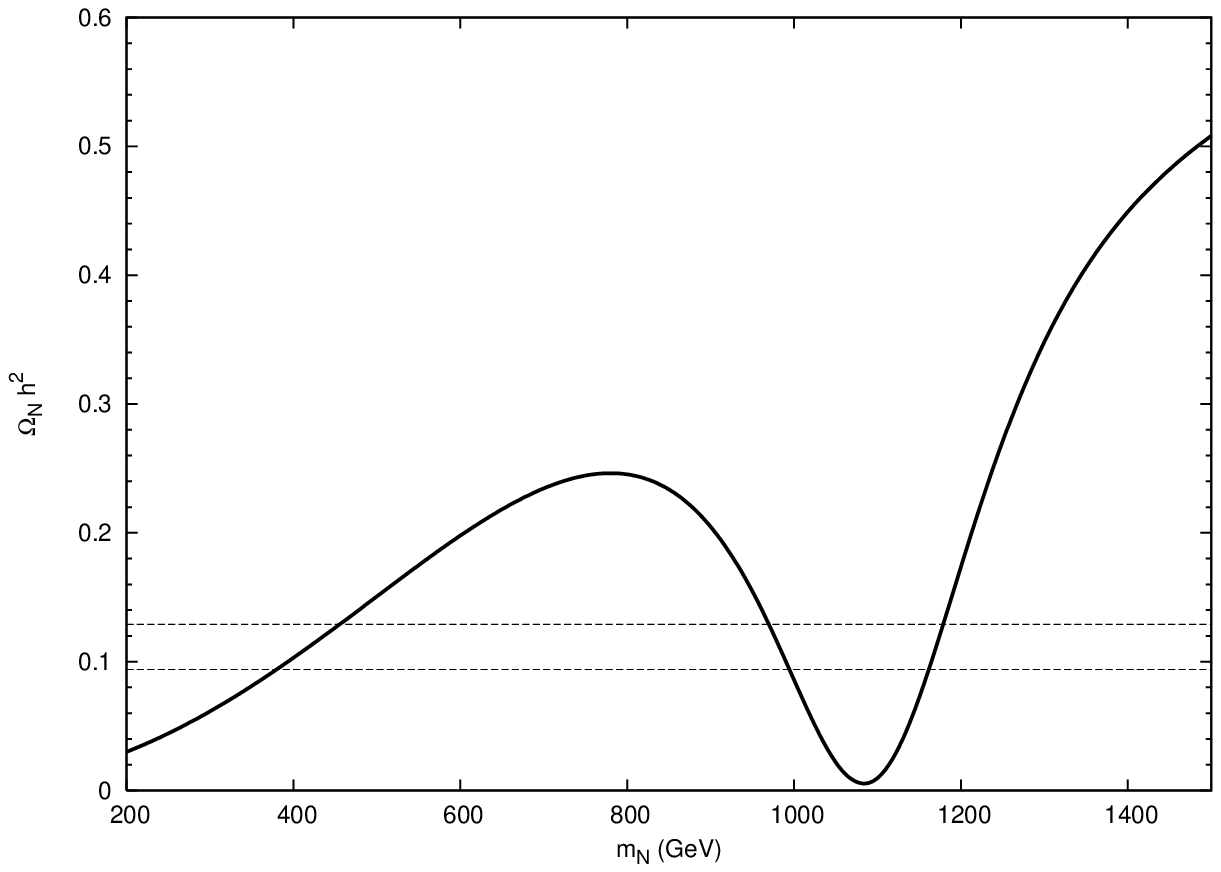}
    \caption{Left: Regions of cosmologically allowed $\Omega_N h^2$ including coannihilation
      with a heavy down type quark. Three different quark masses are shown; $m_D/m_N = 1.1$
      (gray, green online), $1.05$ (light gray, yellow online), degenerate
      (solid black). See figure
      (\ref{fig:relic_N}) for $\Omega_N h^2$ without coannihilation. As before the allowed range for
      $\Omega_N h^2$ is $0.094 \le
\Omega_N h^2 \le 0.129$. Right: $\Omega_N h^2$ versus $m_N$ including
coannihilation with a degenerate heavy quark for $f = 4.0\ \tev$. The dashed
lines indicate the WMAP limits.}
    \label{fig:relic_coD}
  \end{center}
\end{figure}
If the heavy quark is just slightly heavier than the heavy
neutrino, we see that there can be an additional neutrino mass region where heavy neutrino DM is cosmologically
allowed. The range of $m_D/m_N$ where this second region occurs is very
small. For $m_D/m_N > 1.05$, this additional allowed mass region either does not exist or
exists where $m_D$ is experimentally forbidden. Even if $1.0 < m_D/m_N <
1.05$ , the additional region is only acceptable when $f \simge 2.5\ \tev$. For lower $f$,
heavy quark annihilation through gluons is so dominant that $\sigma_{eff}$ is
increased to the point where there is insufficient DM. \\

{\em Coannihilation of $\eta$ with a heavy fermion:} \\

Following the same procedure as above, we can calculate $\Omega_{\eta} h^2$ including coannihilation with a heavy fermion (either heavy
neutrino or heavy quark). The coannihilation channels $\eta N \rightarrow X
X'\ {\rm and}\ \eta D \rightarrow X X'$ are mediated by heavy neutral $W'$ exchange.
The heavy fermions have larger annihilation cross sections and more degrees
of freedom than $\eta$, so their self-annihilation controls $\sigma_{eff}$.

To show the effect of a degenerate heavy fermion on $\Omega_{\eta} h^2$, we plot $\Omega_{\eta} h^2$ including
coannihilation with a heavy neutrino in Figure 5. \\
\begin{figure}[!h]
  \begin{center}
    \includegraphics[width=3.75in,height=2.85in, angle=270]{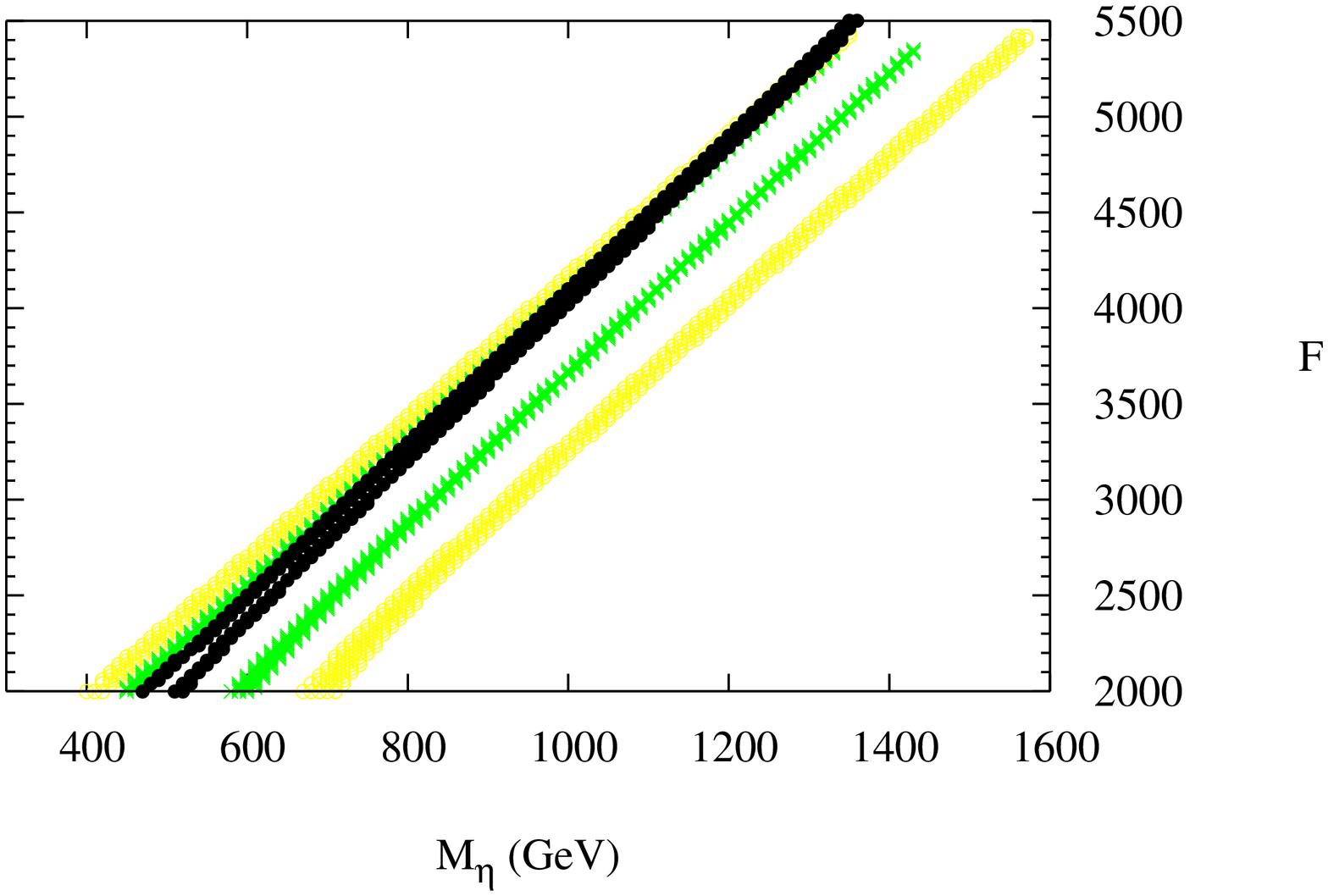}
    \caption{Regions of acceptable $\Omega_{\eta} h^2$ including coannihilation
      with a heavy down neutrino, versus $m_{\eta}$, $f$. Three different
      $N-\eta$ mass differences are shown; $\Delta m =m_N - m_{\eta} = 10\%\
      m_N$ (black), $5\%\ m_N$ (gray, green online), degenerate (light
      gray, yellow online). There is an
      additional region not shown here because it requires $m_{\eta} >
      M_{W'}$. The allowed range for
      $\Omega_N h^2$ is $0.094 \le
\Omega_N h^2 \le 0.129$.}
    \label{fig:relic_coetaN}
  \end{center}
\end{figure}
The most important consequence of coannihilation in this scenario is that the
$s$ channel $Z'$ exchange terms originally in the neutrino self annihilation
cross section are
folded into to the effective cross section for $\eta$.  As a result there is now a large dip in $\Omega_{\eta} h^2$ near $M_{Z'}/2$. For
$m_N$ within $10\% $ of $m_{\eta}$ this dip is
substantial enough that we achieve acceptable $\Omega_N h^2$  for much
lighter $m_{\eta}$. Comparing figures (\ref{fig:relic_eta}) and
(\ref{fig:relic_coetaN}), we can see that coannihilation
decreases the allowed $m_{\eta}$ by almost a factor of $10$. If $m_N$ is even closer to
$m_{\eta}$ the dip becomes larger and we get the correct relic abundance at lighter
$m_{\eta}$. 
 
We now summarize the results of our relic abundance calculations before
investigating possible experimental signatures. In order for $\Omega_{\eta} h^2$ to be consistent with the region
allowed by WMAP for an $\eta$ that is actually the LTOP, we must have coannihilation with a nearly degenerate
fermion. Coannihilating with a heavy neutrino with $m_N/m_{\eta} = 1.1$ for
$f = 4.0\ \tev$, we achieved the
correct relic abundance for $m_{\eta} \sim 1.0\ \tev$. If the LTOP is a heavy neutrino, $\Omega_N h^2$ falls within the
experimental limits for a much wider range of $m_N$ and $f$. Taking $f$ fixed
and varying $m_N$, there are two regions of acceptable $\Omega_N h^2$ - one on
each side of the pole in the cross section at $m_N = M_{Z'}/2$. For $f = 4.0\ \tev$
the two allowed neutrino mass regions are
$m_N \sim 900\ \gev$ and $m_N \sim\ 1300\ \gev$. If the other heavy neutrinos are nearly degenerate with the
LTOP, the allowed mass regions at a given $f$ are closer to
$M_{Z'}/2$. Including coannihilation with a heavy quark, we can
achieve adequate values for $\Omega_N h^2$ for $m_N$ as low as $200\ \gev$. To
achieve a tolerable $\Omega_N h^2$ at such a low $m_N$, $m_D/m_N$ must fall
in a very specific range.

\section*{VIII. Possible Signatures of SLH Dark Matter}

Now that we have identified a few candidate scenarios for DM in the SLH model
with T parity, we look to see what signatures each predicts.

 By looking for anomalous atomic recoil in systems of very cold atoms, an upper limit can be
placed on the DM-nucleon elastic scattering cross section. Currently the most
stringent limits come using $^{73}{\rm Ge}$ in the CDMS~\cite{Armel-Funkhouser:2005zy} experiment located in the
Soudan mine in Minnesota. CDMS places separate limits on the spin independent (SI) and spin dependent
(SD) interactions of DM with nuclei. Spin independent interactions are
generically proportional to the reduced mass of the DM - nuclei system, while
spin dependent interactions are proportional to the spin of the nuclei. For
larger nuclei ($A \ge 40$) and non-Majorana DM, spin independent interactions are usually orders
of magnitude larger than the spin dependent
interactions~\cite{Jungman:1995df,Kamionkowski:1994dp}.

 Based off of the total DM-nucleon elastic cross section, we can also make an
 estimate of the rate for indirect detection. DM particles are collected over time in a massive stellar
object like the sun and they annihilate with each other. The energetic neutrino
remnants from these DM annihilations reach the earth and can be
seen in high energy neutrino detectors.  This
rate does depend somewhat on the dominant
annihilation mode of the DM and other model dependent
parameters~\cite{Jungman:1995df, Halzen:2005ar}. \\

{\em Detection of heavy neutrino DM}\\

Since the heavy neutrino is a
dirac fermion its coupling to the $Z'$ has $(V-A)$ structure. The vector
portion, when combined with the vector part of a
$q \bar q Z'$ interaction, contributes to the spin independent heavy neutrino-nucleon scattering
cross section. One might expect
this contribution to be small given the weakness of the coupling and
the large mass of the $Z'$. A calculation of the $N-$nucleon
effective cross section shows that this is
not the case. The $Z'$ couples to both components of a weak $SU(2)$ doublet
with equal strength. The effective proton/neutron-$Z'$ couplings are then
approximately three times the size of the quark-$Z'$ coupling. The
coherent interaction of a $Z'$ with the protons and neutrons in a large nucleus can
thus be sizable~\footnote{There is also a small contribution from $Z$ exchange, which we include in all
calculations.}. Following~\cite{Griest:1988ma,Drees:1992rr,Drees:1993bu,
Jungman:1995df,Servant:2002hb} we calculate the neutrino-nucleus scattering cross
section (per nucleon) in $~^{73}{\rm Ge}$ to be
\be
\label{eq:Nnucsigma}
\sigma_{N-nuc} \cong 3.7\times10^{-42} \Big(\frac{2\ \tev}{f}\Big)^4\ {\rm
  cm}^2.
\ee
Normalized according to the convention in~\cite{Akerib:2005kh,Brink:2005ej}, the cross
section is nearly independent of the heavy neutrino mass. The only $m_N$
dependence is in the nuclear form factors, which we will ignore here. The
cross section (\ref{eq:Nnucsigma}) therefore applies to both neutrino mass
regions for a given $f$. For
$f \simle 3.0\ \tev$, the $\sigma_{N-nuc}$ from (\ref{eq:Nnucsigma}) is excluded by the current CDMS
data~\cite{Akerib:2005kh, Armel-Funkhouser:2005zy}. Between $f = 3.0\ \tev$ and $f = 4.5\ \tev$,
$\sigma_{N-nuc}$ is beneath the current CDMS limits, but within the predicted
final sensitivities of the current run, CDMSII. Heavy neutrino DM in this
region would certainly be visible in later phases of CDMS, if not earlier. In Fig.6 we show how the
$N$-nucleon scattering cross section per nucleon compares with the current
and predicted CDMS sensitivities~\cite{dmtools} as a function of $f$. \\

\begin{figure}[!ht]
  \begin{center}
    \includegraphics[width=3.0in,height=2.25in]{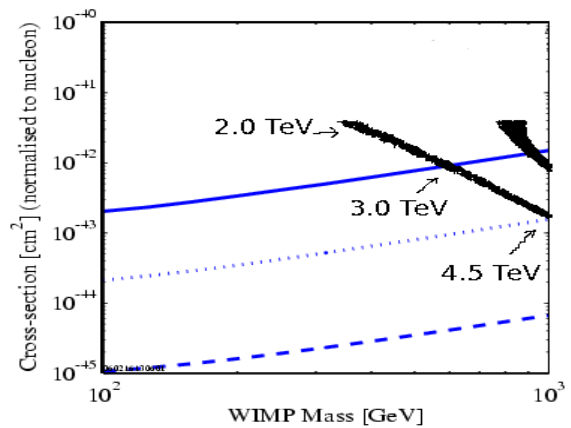}
    \caption{The solid line indicates the current CDMS limit on the spin
      independent DM-nucleon cross section per nucleon as a function of the
      mass of the DM particle. In the shaded region we shown $\sigma_{N-nuc}$ wherever $\Omega_N h^2$ is
      cosmologically allowed, starting with $f=2.0\ \tev$. We assume $m_N
      \ll m_D, m_{\eta}$. As $f$ is
      increased we move down and to the right within the shaded region. The dotted line is the
      projected final sensitivity for CDMS-II. The
      dashed line shows the
      expected sensitivity of the next phase of CDMS, SuperCDMS
      A~\cite{Brink:2005ej}.}
    \label{fig:CDMS}
  \end{center}
\end{figure}

There is also a SD neutrino-nucleus cross
section from the axial-vector portion of $Z'$ exchange. We estimate
it to be several orders of magnitude smaller than the SI cross section and
will therefore ignore its effects. For a
total $(SI + SD)$ nuclear cross section of $\CO(10^{-6}\
pb)$ (per nucleon), we expect $\CO(1)$ event/year per ${\rm km^2}$~\cite{Halzen:2005ar} in
a neutrino telescope.\\

{\em Detection of $\eta$ DM} \\

By the same procedure, we can calculate the effective $\eta-$nucleon cross
section. Since $\eta$ is a scalar, there is no spin dependent interaction.
The full $\eta$-nucleon cross section is just the
spin-independent $\eta$-nucleon cross section.

 An $\eta$ scatters off of a light quark ($u,d,s$) predominantly through
$t$ channel higgs exchange.  Other $\eta q \rightarrow \eta q$ processes, either from $s$ channel
heavy quark exchange or from higher order Yukawa terms like
$\eta^2 (q^{\dag}q)$, are small. The higgs exchange interactions are proportional
$\frac{m^2_{\eta}}{f^2}$ and to the quark masses, thus the strange quark
contribution is the largest.  To get the full SI $\eta$-nucleon cross section, we must also include
$\eta$-gluon scattering $\sigma(\eta g \rightarrow \eta g)$. This occurs
through higgs exchange with a quark loop which emits two gluons. Loops containing
heavy SM quarks ($c,b,t$) or new heavy quarks ($S,D,T$) can make substantial
contributions to $\sigma(\eta g \rightarrow \eta g)$ since
both $\alpha_s$ and the heavy quark Yukawa couplings are
large~\cite{Drees:1992rr,Drees:1993bu,Jungman:1995df}. Using a standard
approximation for the heavy quark (both SM and T-odd) loops~\cite{Shifman:1978zn} we
calculate the $\eta-$nucleon cross section in the limit $m_{T},
m_{D} \gg m_{\eta}$ to be
\be
\label{eq:etanucsig}
\sigma_{\eta-nuc.} \sim 10^{-45} \Big(\frac{m_{\eta}}{500\ \gev} \Big)^2
\Big(\frac{2.0\ \tev}{f} \Big)^4\ {\rm cm^2}.
\ee
This cross section is well below the current and projected CDMS
limits for the entire $m_{\eta}-f$ range of interest. Given $\sigma_{\eta-nuc} \simle 10^{-9}\ {\rm pb}$ and the lack of any spin
dependent interactions with the nucleus, the potential for indirect
detection of $\eta$ in a neutrino telescope is also very
dim~\cite{Halzen:2005ar}. \\

\section*{IX. Conclusion}

One consequence of enforcing a $\CZ_2$ symmetry onto a Little
Higgs model is that the lightest odd particle is a potential DM
candidate. We have investigated this here in the context of the Simplest Little Higgs
model. We are unable to make all new particles odd, and therefore we are
forced to work within a somewhat finely tuned model. Accepting this fine
tuning, we examined the two DM candidates in this model; a heavy neutrino $(N,n^c)$ and a scalar
$\eta$. Through the standard relic abundance calculations we have found the
circumstances under which these DM candidates are allowed by cosmology. We
have also checked to make sure these circumstances are consistent with current
direct detection limits.

The first DM candidate we investigated was the scalar $\eta$.  We found that $\eta$ cannot be the LTOP unless there is a nearly degenerate
(within $10\% $) fermion. Coannihilating with a nearly degenerate heavy fermion, there is
parameter space where $\Omega_{\eta} h^2$ is within the WMAP limits and
$m_{\eta}$ is in the $\tev$ range. For $f = 4.0\ \tev$, this occurs for
$m_{\eta} \sim 1.0\ \tev$, although the actual number will depend on the type
of fermion and the degree of degeneracy. We estimate 
the $\eta - nucleon$ cross section to be $ \simle \CO(10^{-9})\ pb$ in the
entire region of interest. This cross
section is well below the
predicted sensitivity bounds of both the current CDMS run and the first
SuperCDMS phase. $\eta$ DM in this scenario would be difficult to find.

The other DM candidate we considered was a heavy neutrino. When the mass
of the neutrino is close $M_{Z'}/2$, the annihilation cross section is
enhanced and lowers $\Omega_N h^2$ into an acceptable range. Typically this
happens for $|m_N - M_{Z'}/2| \sim 200\ \gev$. Provided that the neutrino
mass is in the right range, $\Omega_N h^2$ stays in the allowed region for a large
range of the other T-odd particle masses. However, heavy neutrino DM is ruled out by direct detection
unless $f \ge 3.0\ \tev$. Based on the projected final CDMS II sensitivity (see fig.(\ref{fig:CDMS})), the bound on $f$
could become as high as $4.5\ \tev$ if no heavy neutrino DM is seen by the end of
the current CDMS run. These bounds are approximately $m_N$ independent,
although CDMS constraints on more massive particles are somewhat weaker. The
bound $f \simge 4.5$ is consistent with the strictest EWPO
bound~\cite{Marandella:2005wd, Han:2005dz}. Heavy neutrino DM for $f = 4.5\ \tev$ has mass $\sim 1\ \tev$, although it can be much
lighter if a heavy quark is nearly degenerate with the heavy neutrino. As neither the heavy neutrino nor $\eta$ has 
significant SD interactions with nuclei, we expect 
indirect detection signals of SLH DM to be very small.

Some other distinct features of this model are the T-even $Z'$, a very heavy
$m_T \sim f$ top quark partner, and the lack of
heavy T parity even quarks~\cite{Cheng:2004yc,Hubisz:2004ft}. In addition to
the direct detection signals we discussed, this model and
its DM candidates may yield interesting collider phenomenology which
remains to be investigated.

 \section*{Acknowledgments}

 First and foremost we would like to thank Martin Schmaltz for his help and guidance
 throughout this work. We are also grateful to Tuhin Roy for many helpful discussions
 and for reading early drafts. This work was supported by the U.S. Department
 of Energy under Grant No. DE-FG02-91ER40676.





\bibliography{SLH.bib}
\bibliographystyle{utcaps}
\end{document}